# Uniform Polynomial Equations Providing Higher-order Multi-dimensional Models in Lattice Boltzmann Theory

Jae Wan Shim

*KIST and the University of Science and Technology, 136-791, Seoul, Korea*

**Abstract.** We present a set of polynomial equations that provides models of the lattice Boltzmann theory for any required level of accuracy and for any dimensional space in a general form. We explicitly derive two- and three-dimensional models applicable to describe thermal compressible flows of the level of the Navier-Stokes equations.

Since Ulam and von Neumann's concept of the cellular automata, where the collective behavior of the entire system constituted of each automaton governed by a simple rule that the finite number of states of each automaton at time $t_{n+1}$ is determined by that of its neighbors at time $t_n$ could be sufficiently complex under certain conditions, the applications of the idea were developed in many areas [1]. In the field of fluid flows, Frisch, Hasslacher, and Pomeau [2] introduced the lattice-gas automata as a model of the Navier-Stokes equations, which led to the development of the lattice Boltzmann theory [3]. Recently, various attempts have been performed to derive higher-order accuracy models including the works of [4-6]. However, their frameworks are insufficient to derive higher-order accurate and efficient models in systematical ways.

Here, we present a set of polynomial equations in a single form. This enables us to obtain multi-dimensional models having accuracy up to any required level even beyond the Navier-Stokes equations by a systematical way for the purpose of increasing efficiency by reducing the number of discrete velocities with keeping accuracy. We explicitly derive efficient multi-dimensional models having smaller number of discrete velocities than the previously known.

Fluid flows described by the lattice Boltzmann equation use the notion of fictitious particles moving their positions and changing their velocity distribution according to a simple rule $f_i(\mathbf{x}+\mathbf{v}_i,t+\Delta t) = (1-\omega)f_i(\mathbf{x},t) + \omega f_i^{eq}(\mathbf{x},t)$ where $f_i(\mathbf{x},t)$ is the density of particles having discrete velocities $\mathbf{v}_i$ at position $\mathbf{x}$ at time $t$. The density distribution $f_i^{eq}(\mathbf{x},t)$ is the equilibrium state of $f_i(\mathbf{x},t)$ and $\omega$ adjusts viscosity. For the continuous velocity space which is not in this case, the equilibrium state is the Maxwell-Boltzmann distribution $f^{eq} = \rho(\pi\theta)^{-D/2}\exp(-\|\mathbf{v}-\mathbf{u}\|^2)$ where $\rho$ is the density, $D$ the dimension of space, and the dimensionless variables defined by $\theta \equiv 2k_B T/m_g$, $\mathbf{v} \equiv \theta^{-1/2}\mathbf{V}$, and $\mathbf{u} \equiv \theta^{-1/2}\mathbf{U}$. The symbol $T$ is the temperature, $k_B$ the Boltzmann constant, $m_g$ the molecular mass, $\mathbf{V}$ the microscopic (or molecular) velocity, and $\mathbf{U}$ the macroscopic velocity that is a local average of the microscopic velocity. For the discrete velocity space, we use the following constraints to obtain $f_i^{eq}(\mathbf{x},t)$;

$$\mathbf{M}(n) \equiv \int \mathbf{v}^n f^{eq}(\mathbf{v}) d\mathbf{V} = \sum_i \mathbf{v}_i^n f_i^{eq}(\mathbf{v}_i) \text{ for } n = 0,1,...,m \quad (1)$$

where $\mathbf{v}^n \equiv v_{x_1}^{a_1} v_{x_2}^{a_2} ... v_{x_D}^{a_D}$ and $\mathbf{v}_i^n \equiv v_{i,x_1}^{a_1} v_{i,x_2}^{a_2} ... v_{i,x_D}^{a_D}$ such that $a_1 + a_2 + ... + a_D = n$ and $a_\alpha \in \mathbb{N}_0$ for $\alpha = 1,2,...,D$. The subscripts $x_i$ for $i = 1,2,...,D$ signify the coordinates in $D$-dimensional Cartesian coordinate system. In Formula (1), the maximum order $m$ of the momentum $\mathbf{M}$ is one of the factors determining the accuracy of the discrete models to be obtained. The constraints satisfy the conservation of physical properties such as mass, momentum, pressure tensor, energy flux, and the change rate of the energy flux *etc.* as the order of $\mathbf{M}$ increases. If we express $f^{eq}$ by series expansions such as the Taylor and the Hermite having the form of $f^{eq}(\mathbf{v}) \approx \exp(-v^2)P(\mathbf{v})$ where $P(\mathbf{v})$ is a polynomial in $\mathbf{v}$ and $v^2 \equiv \mathbf{v}\cdot\mathbf{v}$, we can obtain $f_i^{eq}$ in the form of

$$f_i^{eq}(\mathbf{v}_i) = w_i P(\mathbf{v}_i) \quad (2)$$

where $w_i$ are constant weight coefficients. With the use of Formula (2), Formula (1) is satisfied up to the order of the velocity momentum $m$ for any $P(\mathbf{v}_i)$ of degree $k$, if and only if

$$\int \mathbf{v}^n \exp(-v^2) d\mathbf{v} = \sum_i w_i \mathbf{v}_i^n \text{ for } n = 0,1,...,m+k. \quad (3)$$

The left side of Formula (3) can be evaluated by

$$\int \mathbf{v}^n \exp(-v^2) d\mathbf{v} = \begin{cases} \prod_{\alpha=1}^{D} \Gamma((a_\alpha+1)/2) & \text{for even } a_\alpha, \\ 0 & \text{otherwise} \end{cases} \quad (4)$$

where $\Gamma$ is the Gaussian Gamma function which can be expressed by the double factorial as $\Gamma((n+1)/2) = \sqrt{\pi}(n-1)!!/2^{n/2}$. If we assume that the discrete equilibrium distribution $f_i^{eq}$ is isotropic when $\mathbf{u} = \mathbf{0}$ like the Maxwell-Boltzmann distribution, Formula (3) is satisfied when $n$ is odd. Therefore, Formula (3) can be written by

$$\sum_i w_i \mathbf{v}_i^n = \prod_{\alpha=1}^{D} \Gamma((a_\alpha+1)/2) \text{ for } n = 0, 2, 4, ..., n_{max} \leq m+k \text{ and even } a_\alpha. \quad (5)$$

Note that $m$ and $k$ are two factors determining the accuracy of the discrete models. If we assume that we expand the Maxwell-Boltzmann distribution by the Hermite expansion for Formula (2), the degree $k$ of $P(\mathbf{v})$ should be $m$ at least to satisfy up to the $m$-th order momentum $\mathbf{M}(m)$ in Formula (1). Therefore, the weight coefficients $w_i$ and the discrete velocities $\mathbf{v}_i$ should satisfy Formula (5) up to $n_{max} = 2m$. Let us define $\pi(q, D)$ by the number of possible ways of representing $q$ as a sum of natural numbers such that the number of terms are equal or less than $D$ and we additionally define $\pi(0, D) = 1$. Then, the number of equations provided from Formula (5) up to $n_{max} = 2m$ can be calculated by $\sum_{q=0}^{m} \pi(q, D)$.

For a thermal compressible flow of the level of the Navier-Stokes equations, Formula (1) should be satisfied up to $m = 4$. Then, the numbers of equations for two- and three-dimensional spaces are $\sum_{\bar{n}=0}^{4} \pi(\bar{n}, 2) = 9$ and $\sum_{\bar{n}=0}^{4} \pi(\bar{n}, 3) = 11$, respectively. In the case of two-dimensional space, we have 9 equations to satisfy, therefore, we make models having 9 unknown variables as in Table 1. To satisfy isotropic condition, we generate discrete velocities from the representative ones by the symmetries about $y$-axis, $x$-axis, and $y = x$ line in the Cartesian coordinate system. Then, we obtain the two-dimensional 33-velocities model. For three-dimensional models, we present an example, the 95-velocities model in Table 2. Note that the models derived here have smaller sets of discrete velocities than the previously known models such as the two-dimensional 37-velocities model and the three-dimensional 107-velocities model [4] without losing accuracy. Of course, we can obtain the low-order M(2) accuracy models such as the three-dimensional 15- and 19-velocities ones as well [7].

| group | number of discrete velocities | representative discrete velocity ($c = .819381$) | weight coefficient |
|---|---|---|---|
| 1 | 1 | $(0,0)$ | .161987 |
| 2 | 4 | $c(1,0)$ | .143204 |
| 3 | 4 | $c(1,1)$ | .0338840 |
| 4 | 4 | $c(2,0)$ | .00556112 |
| 5 | 4 | $c(2,2)$ | $8.44799 \times 10^{-5}$ |
| 6 | 4 | $c(3,0)$ | .00113254 |
| 7 | 8 | $c(2,1)$ | .0128169 |
| 8 | 4 | $c(4,4)$ | $3.45552 \times 10^{-6}$ |

Table 1. Detail of the two-dimensional 33-velocities model.

| group | number of discrete velocities | representative discrete velocity ($c = .421803$) | weight coefficient |
|---|---|---|---|
| 1 | 1 | $(0,0,0)$ | .206847 |
| 2 | 6 | $c(2,0,0)$ | .00442257 |
| 3 | 12 | $c(2,2,0)$ | .0333341 |
| 4 | 8 | $c(2,2,2)$ | .0128902 |
| 5 | 6 | $c(3,0,0)$ | .0287920 |
| 6 | 12 | $c(3,3,0)$ | .00264319 |
| 7 | 8 | $c(3,3,3)$ | .000927908 |
| 8 | 24 | $c(2,2,5)$ | .00106078 |
| 9 | 12 | $c(4,4,0)$ | .000804376 |
| 10 | 6 | $c(5,0,0)$ | .00274697 |

Table 2. Detail of the three-dimensional 95-velocities model.

We demonstrate the accuracy and the stability of the three-dimensional 95-velocities model by the shock tube problem. The dimension of the nodes of the shock tube is $X \times Y \times Z = 11 \times 11 \times 800$. The two homogeneous initial conditions $C_L = \{\rho = p = 4, \theta = 1, \mathbf{u} = \mathbf{0}\}$ and $C_R = \{\rho = p = 1, \theta = 1, \mathbf{u} = \mathbf{0}\}$ are applied to the left ($1 \le Z \le 400$) and the right ($401 \le Z \le 800$) half spaces. The boundary conditions on the $XZ$- and $YZ$-planes are periodic and those on the left and the right $XY$-planes are $C_L$ and $C_R$, respectively. Figure 1 shows the density and the temperature profiles along the central longitudinal axis of the shock tube. The excellent agreements of the values between the analytical solutions and the simulation results are observed on the plateaus. Note that the result of the shock tube simulation with the model having the third-order momentum accuracy shows slight mismatches with respect to the analytical solution on the plateaus [8]. The difference of the steepness on the shock front comes from the difference of the viscosity. The analytical solution of the Riemann problem, which deals with the zero viscosity flow, is steeper than the result of the simulation dealing with the viscous flow with $\omega = 1.5$.

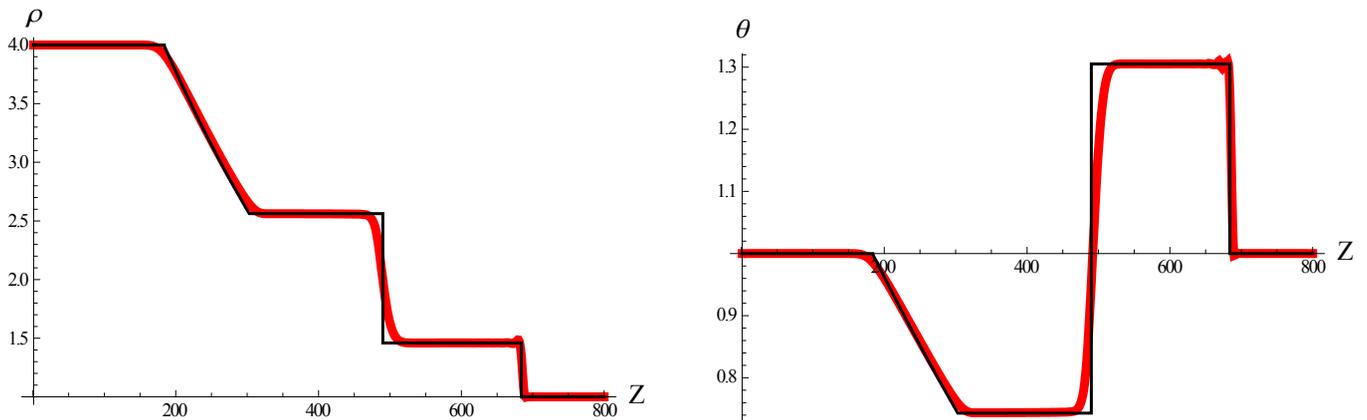

**FIGURE 1.** The density and the temperature profiles of the shock tube simulation are shown on the left and the right, respectively. The analytical solution (thin black line) of the Riemann problem and the simulation results of $\omega = 1.5$ (thick red line) are drawn.

We conclude this paper with a remark that we can obtain the models having higher-order momentum accuracy for the purpose of the level of accuracy of the Burnett or the super Burnett equations from the set of polynomial equations Formula (5) expressed in a general form.

**ACKNOWLEDGMENTS**

This work was partially supported by the KIST Institutional Program.